\documentclass[11pt]{article}
\RequirePackage{cite}
\RequirePackage{times}
\RequirePackage{fullpage}
\RequirePackage{ifthen}
\RequirePackage{graphicx}
\RequirePackage{color}
\definecolor{gray}{gray}{0.35}
\begin{document}
\def\@cite#1#2{$^{\mbox{\scriptsize #1\if@tempswa , #2\fi}}$}
\setlength{\parindent}{0in}
\setlength{\parskip}{8pt}

\renewenvironment{abstract}{%
    \setlength{\parindent}{0in}%
    \setlength{\parskip}{0in}%
    \bfseries%
    }{\par\vspace{-6pt}}
\renewcommand\refname{\vspace{-48pt}\setlength{\parskip}{8pt}}
\begin{flushleft}
{\Large
\textbf{Excitable human dynamics driven by extrinsic events in massive communities}
}

{Joachim Mathiesen$^{1,\ast}$, Luiza Angheluta$^2$, Peter T.~H. Ahlgren$^1$ and Mogens H. Jensen$^1$}\\
$^1${Niels Bohr Institute, University of Copenhagen, Copenhagen, Denmark}\\
$^2$ {Department of Physics, Physics of Geological Processes, University of Oslo, Oslo, Norway}\\
$\ast$ E-mail: mathies@nbi.dk
\end{flushleft}

\begin{abstract}
Using empirical data from a social media site (Twitter) and on trading volumes of financial securities, we analyze the correlated human activity in massive social organizations. The activity, typically excited by real-world events and measured by the occurrence rate of international brand names and trading volumes, is characterized by intermittent fluctuations with bursts of high activity separated by quiescent periods. 
These fluctuations are broadly distributed with an inverse cubic tail and have long-range temporal correlations with a $1/f$ power spectrum. We describe the activity by a stochastic point process and derive the distribution of activity levels from the corresponding stochastic differential equation. The distribution and the corresponding power spectrum are fully consistent with the empirical observations. 
\end{abstract}

\section*{Introduction} Online social networks have emphatically changed the way people interact. The development of network theories and growth in available data on human behaviour~\cite{king2010,Ginsberg2008,Borgatti2010,DarwinEinstein2005} have prompted an explosive interest in research on the evolution of behaviors~\cite{Bul2010,RBHLM09, C10} and social structures~\cite{ADDG04, BGLL08}. 
Among the many forms of online social media, microblogging services, such as Twitter~\cite{KLPM10, Twitter2011, Huberman2009}, are characterized by a real-time dynamics with large numbers of user broadcasts related to real-world events. Twitter is a popular microblogging platform where a registered user can submit small pieces of information, named ``tweets'', that are either private or made public to the user's followers. The length of a tweet is limited to 140 characters and its content ranges widely from personal updates to massively distributed advertisements or political messages. Twitter has a global outreach and, hence, is used by an increasing number of companies and political organizations to disseminate news. To a large extent, Twitter users can be seen as direct social sensors to measure the popularity of various topics. A large part of recent research on Twitter utilizes user activity as predictor for real world events including the dynamics of stock-market prices~\cite{BMZ11}, box-office revenues~\cite{AH10}, real-time detection of the location of earthquakes hitting populated areas~\cite{Sakaki10} and for opinion mining and political sentiment analysis~\cite{TSSW10}. Large-scale behavioral data from other online media have been shown to have a similar predicting power, e.g. Google query volumes, have been used to detect early signs of stock market moves~\cite{PMS13} or more general movements in society~\cite{CV12}. 

\begin{figure}
\begin{center}
\includegraphics[width=.85\textwidth]{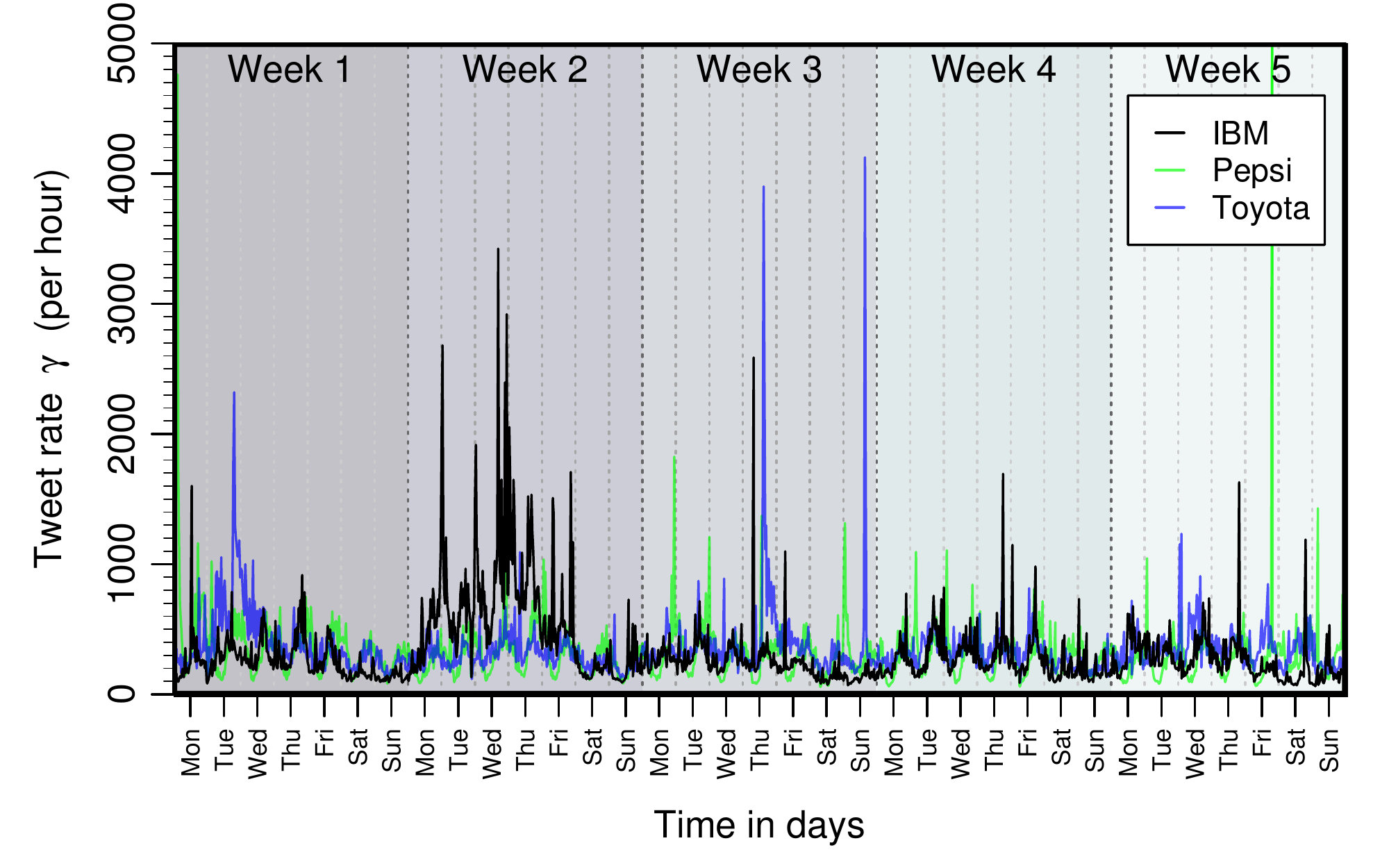}
\caption{Temporal variation of tweet rates of three international brands, IBM, Pepsi and Toyota. The time signals are for all brands intermittent, i.e.~they have longer periods of relatively steady activity levels interrupted by sudden high activity spikes. The time signals are all modulated by an underlying periodic variation over days and weeks.}
\label{fig1}
\end{center}
\end{figure}

While behavioral data from Twitter have been suggested to predict many real world events or have been used in mapping out social networks, the statistics of the combined user activity are not well understood. Here we suggest a stochastic model for the user activity in massive online communities. Our model sheds light on the statistical properties of the large-scale user activity on Twitter as well as the underlying correlations. Similar to the user activity on Twitter, trading volumes on the stock market reflect the interest that investors have in particular securities or products at given prices. Interestingly, as we will point out below, we find that the trading activity on financial securities is quite similar to the large-scale user activity on Twitter.

We have automatically queried Twitter for tweets containing one or more international brand names. These tweets appear at highly irregular time intervals, see Fig.~1, reflecting intermittent user activity levels. We consider the broadcasting of tweets to be a random point process with large fluctuations in the time intervals between the online appearances of messages. The number of tweets containing a certain brand, "A", as a function of $t$ is therefore given by a time signal $g_A(t)$ composed of isolated events occurring at random times $t_\ell$, 
%
$g_A(t)=\sum_\ell \delta(t-t_\ell), $
%
where the index $\ell$ refers to a specific posting event. For each query, a number of the latest tweets $n_A\le 1500$ is returned. Thus, we determine an average tweet rate for a given query $k$ as
\begin{equation}\label{eq:gamma_A}
\gamma_A(t_{n_A})=\frac{1}{t_{n_A}-t_1} \int_{t_1}^{t_{n_A}} \mathrm{d} t g_A(t) =\frac{n_A(t_{n_A})}{t_{n_A}-t_1},  
\end{equation}
where $t_1$ and $t_{n_A}$ correspond to the time of the oldest, respectively latest tweet returned by the query $k$. The time interval $\tau_k=t_{n_A}-t_1$ for a fixed number $n_A$ of tweets is a highly fluctuating variable from one query to another. 
We notice that the appearance of tweets on Twitter resembles a non-homogeneous Poisson process with random fluctuations in the average tweet rate $\gamma_A$.

We have further collected data for the trading volume of selected equities over a period similar to that covered by the Twitter data. In particular, we consider the trading volume in the three shares Apple Inc. (AAPL), Nokia Corporation (NOK), and Green Mountain Coffee Roasters Inc. (GMCR). The number of shares traded for each security was accumulated and sampled in one-minute intervals. Changes in trading volume and price are known to be highly leptokurtic~\cite{Cont00}, have long range correlations~\cite{Bonanno00} and intriguing scaling properties~\cite{stanley,Stanley03}. We consider the volume as a simple proxy for the temporal interest that the market has in a given security disregarding more complex effects that might influence the price formation process. 

In the Twitter data, we distinguish two types of user activity, one where users post messages independently of other tweets and one where users interact directly, e.g.~ by re-posting information from other users in so-called retweets or by submitting responses to existing tweets. In general, a retweet contains text from the original tweet together with a reference to the author who posted it. Retweets typically form a smaller subset of all tweets and the frequency by which individual tweets are retweeted follows a power law with an exponent similar to the out-degree distribution of Twitter users. The out-degree is here measured by the number of followers of a user, i.e. the number of people that receives directly tweets posted by that user. This suggests that the rate of information spread by re-posting goes proportional with the number of followers and is limited by the local network topology. However, since there are only few tweets that generate a large flux of retweets, this may not be the most efficient way to diffuse information between Twitter's users. There are information pathways that are not only related to the network topology, but, to a larger degree, correspond to many users tweeting at the same time about the same thing triggered by events outside the network.

The intermittent dynamics of individuals has previously been modelled in terms of a timing selection mechanism between different tasks~\cite{barabasi05b}. The prioritization of various tasks is suggested to lead to a bursty dynamics with power-law distributed waiting times. This is in contrast to a homogeneous Poisson process, where the waiting time between tasks that are being selected at random follows an exponential distribution. %
Here, we propose a global measure of collective human behavior and introduce a stochastic model for the global activity rate associated with many interacting individuals in a large social organization. The activity rate is characterized by long-term memory effects as well as non-exponential distributed waiting times.

\begin{figure}
\centering
\includegraphics[width=.45\textwidth]{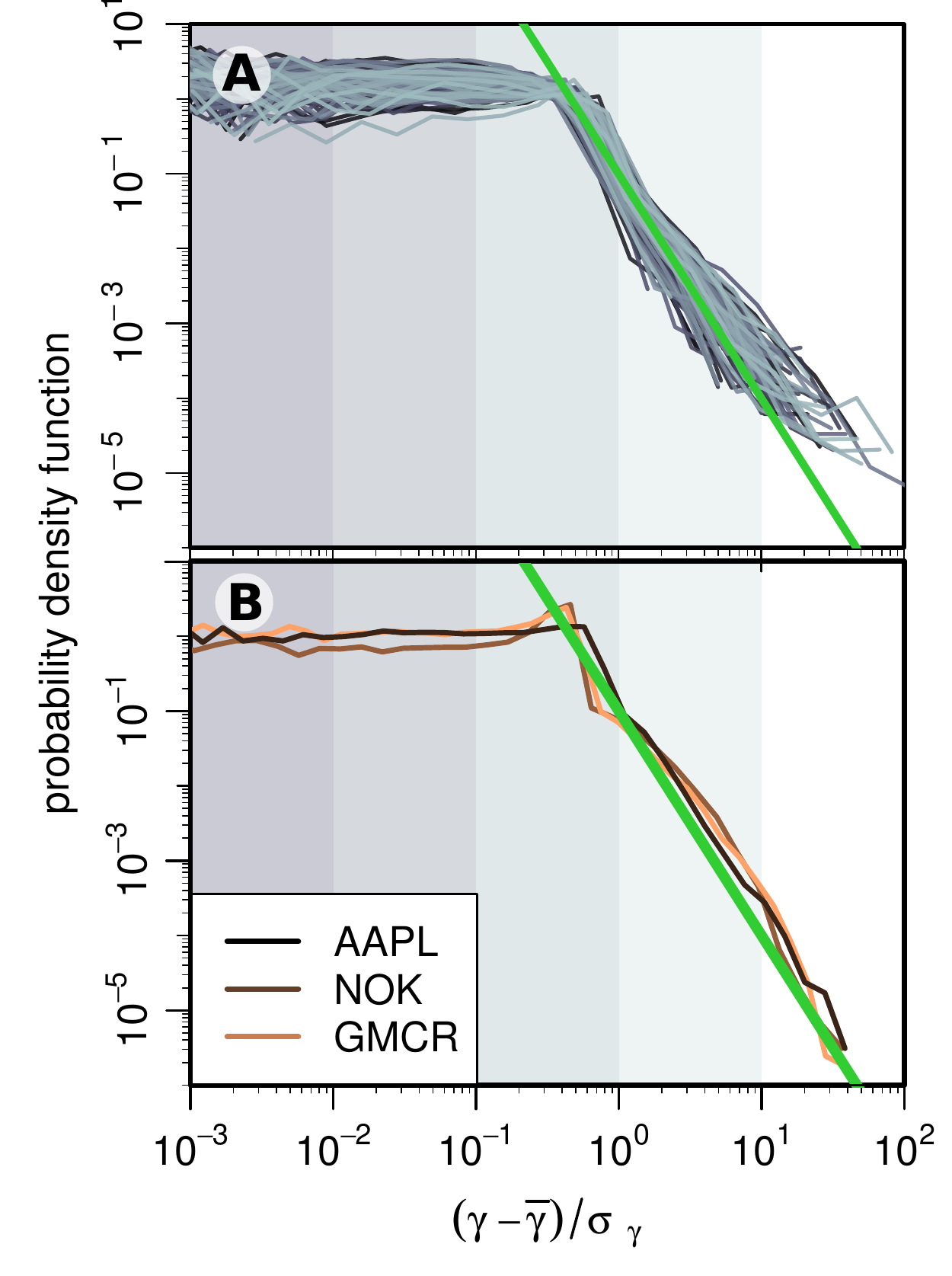}%
\caption{Probability density functions of A) brand tweet rates and B) trading volume rates for stocks. The mean values have been subtracted from the individual rates and the rates have been normalized by their respective standard deviations. The density functions collapse and reveal a common scaling behaviour for relatively large rates. The green line added to both panels is a guide to the eye and is consistent with a scaling exponent of $-3$. In panel B, we consider trading activities in the companies Apple Inc.~(AAPL), Nokia Corporation (NOK), and Green Mountain Coffee Roasters Inc. (GMCR).}
\label{fig2}
\end{figure}

\section*{Results}
Interestingly, the intermittent tweet rates of specific brands follow a distribution $P_0(\gamma)$ with a power-law tail with an exponent $\beta$ close to an inverse cube, $\beta=2.9\pm 0.4$ (s.d.), as seen in Fig.~2. For the trading volumes, we achieve values for the exponent $\beta_{\mbox{\tiny AAPL}}=2.9$, $\beta_{\mbox{\tiny NOK}}=3.1$, and $\beta_{\mbox{\tiny GMCR}}=3.0$.  Moreover, the fluctuations in the flux of tweets are long-range correlated with a power spectrum that decays as $1/f^\alpha$, where $f$ is the frequency and $\alpha=1.0\pm 0.4$(s.d.), ind an intermediate frequency window corresponding to timescales from 20 minutes to 24 hours as shown in Fig.~3. This means that tweets posted at a given time are influenced with an equal strength by tweets on all time scales ranging back as far as $\approx 24$ hours. At the same time, high bursts of new tweets, extreme events, occur in the tail of $P_0(\gamma)$. The $1/f$-noise is a widespread phenomenon observed in a variety of different systems, including voltage fluctuations~\cite{Voss76}, heartbeats~\cite{Kobayashi82}, free-way traffic~\cite{Zhang95}, music~\cite{Voss78}, trades in financial securities~\cite{Bonanno00}, along with many other examples. Although, there is not a unified theory that would apply to all systems exhibiting $1/f$-noise, there are numerous models that reproduce the $1/f$-fluctuations in temporal signals with fluctuations drawn from different distributions. On the other hand, there are also plenty of studies that focus on the non-Gaussian, power-law statistics of time signals, $P_0(\gamma)\sim \gamma^{-\beta}$, independent of their power spectrum. More recent studies on stochastic point processes investigated the relationship between Pareto-type distributions of the variables and their $1/f^\alpha$ power spectrum, e.g.~\cite{Kaulakys05,Ruseckas10,Erland11}. The idea behind a stochastic point process is to model the average waiting time between random, discrete events by a multiplicative noise process. Essentially, the complexity of scale-free distributed variables with a $1/f^\alpha$ spectral density emerges from the multiplicative noise. 
Here we present a stochastic point process that captures both the $P_0(\gamma)\sim \gamma^{-3}$ and $1/f$-noise features. Other models of correlated human behavior~\cite{kertesz12} predict similarly a power law distribution of the activity rates. However, these models possess a weaker memory effect and do not reproduce the scaling exponent for the power spectrum that we observe.

We assume a scenario where the human activity in the case of no external input is determined by a natural drift towards inactivity. That is the waiting time $\tau$ since the last activity increases with time $\partial_t \tau=1$. On the other hand, excitation by external events drive the system towards higher activity levels. Without correlations in the user activity, the rate $\gamma(t)$ is determined by a balance between the drift towards inactivity and repeated excitation. We assume that the correlation in the human dynamics is controlled by a current waiting time between events in the shape of a multiplicative noise with an amplitude given by $\sqrt\tau$. The stochastic process for the average waiting time is therefore determined by a stochastic differential equation on the form 
\begin{equation}\label{eq:tau}
 \frac{d\tau}{dt} = 1+f(\tau)+\sqrt{\tau}\eta(t),
\end{equation}
where the $\eta$ is Gaussian noise with zero mean and unit variance and the deterministic part $f(\tau)$ is chosen such that the process attains a non-trivial stationary distribution.
Collective interactions between users sending messages on Twitter is effectively modeled by the intrinsic, multiplicative noise. The amplitude of the intrinsic noise term is proportional to $\sqrt{\tau}$, which implies that, if the dynamics were solely driven by noise, the waiting time $\tau$ would have an absorbing state, i.e. $\tau=0$, corresponding to a tweeting activity that is constant in time and never ceases. The drift term equal to $1$ is added to mimic that, if nothing happens, the average waiting time would increase linearly with time as mentioned above. Due to this constant drift term, the absorbing state $\tau=0$ is never attained, although there are sudden excursions in its neighborhood. The stationary probability distribution function of waiting times, $F(\tau)$ corresponding to Eq.~(\ref{eq:tau}) is obtained as the steady-state solution of the Fokker-Plank equation in the Ito formulation and given as 

\begin{figure}[t!]
\centering
\includegraphics[width=.45\textwidth]{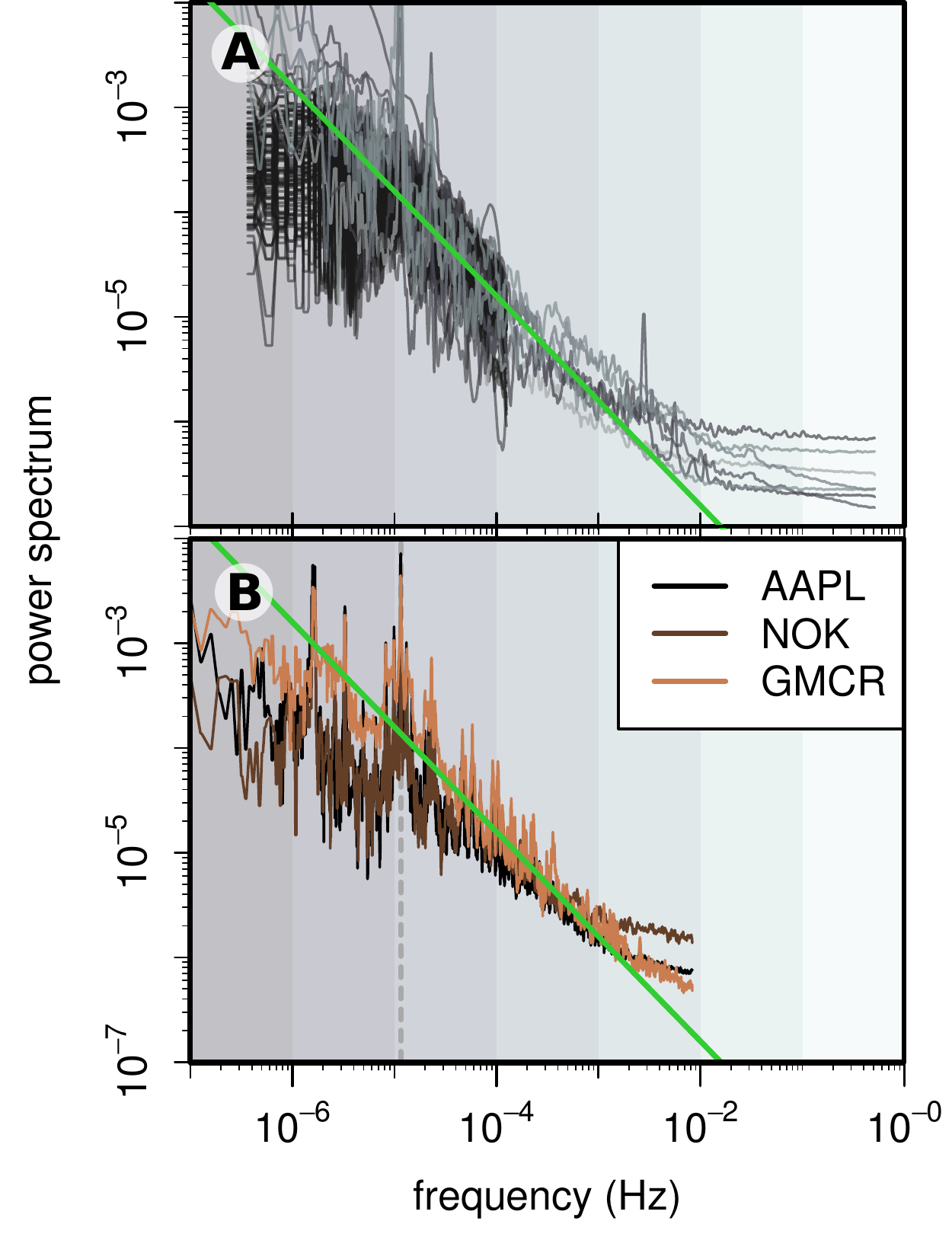}%
\caption{Power spectra of the activity rates $\gamma(t)$ for the individual brands (panel A) and stocks (Panel B). For high frequencies 0.1~-~24 hours$^{-1}$, the spectra have a characteristic $1/f$ behavior with a crossover to white noise at very high frequencies. The green line is a guide to the eye and corresponds to a $1/f$ behavior.}
\label{f1}
\end{figure}

\begin{equation}\label{eq:Ptau}
F(\tau)\sim \tau e^{2\int_0^\tau f(\tau')/\tau' d\tau'},
\end{equation}
apart from the normalization constant. The divergence at large $\tau$'s is suppressed by the cut-off function which depends on $f(\tau)$. However, at small $\tau$'s, corresponding to large tweet rates, $\gamma$, the function $f(\tau)$ is irrelevant since $F(\tau)\sim \tau$. Thus, in the scaling regime, we can safely ignore $f(\tau)$. Using that $\gamma\sim 1/\tau$, the stochastic dynamics for $\gamma$ follows from Eq.~(\ref{eq:tau}) by Ito's lemma and given as 
\begin{equation}\label{eq:gamma}
 \frac{d\gamma}{dt} = \gamma^{3/2}\eta(t),
\end{equation}
where we ignore the contributions due to $f(\tau)$ that only determines the range over which $\gamma(t)$ is power-law distributed as $P_0(\gamma)\sim \gamma^{-3}$. The power spectrum of tweet rate fluctuations is determined by the joint distribution $P(\gamma,t;\gamma',t')$ associated with the stochastic process in Eq.~(\ref{eq:gamma}). 
By the Wiener$-$Khintchine theorem the power spectral density of $\gamma$ is related to the correlation function as
\begin{eqnarray}\label{eq:Sf}
S(f) &=& 4\int_{0}^{\infty} dt \langle\gamma(0)\gamma(t)\rangle \cos(2\pi f t),
\end{eqnarray}
where the correlation function $\langle \gamma(t)\gamma(0)$ is defined as
\begin{equation}\label{eq:corr}
 \langle\gamma(t)\gamma(0)\rangle = \int d\gamma d\gamma' P(\gamma,t;\gamma',0)\gamma \gamma'.
\end{equation}
As with other point processes one may assume that the transition distribution has an eigenfunction expansion, and therefore 
\begin{equation}\label{eq:eignPn}
 P(\gamma,t;\gamma',0) \approx \sum_n  P_n(\gamma)P_n(\gamma')e^{-\lambda_n t}
\end{equation}
where the unnormalized probability eigenfunctions $P_n(\gamma)$ satisfy the master equation corresponding to Eq.~(\ref{eq:gamma_A}) and are given by 
\begin{equation}
 -\lambda_n P_n(\gamma) = \frac{1}{2}\frac{d^2}{d\gamma^2}\left(\gamma^3 P_n(\gamma)\right). 
\end{equation}
Combining Eqs.~(\ref{eq:corr}) and (\ref{eq:eignPn}), we have that 
\begin{equation}
 \langle\gamma(t)\gamma(0)\rangle = \sum_n \gamma_n^2 e^{-\lambda_n t},
\end{equation}
where $\gamma_n = \int_0^\infty d\gamma \gamma P_n(\gamma)$ is the first moment of the probability eigenfunction. By its Fourier transform as in Eq.~(\ref{eq:Sf}), the power spectrum can be written as a sum of Lorentzian spectra
\begin{eqnarray}
S(f) \approx 4\sum_n \frac{\gamma_n^2 \lambda_n}{\lambda_n^2+4\pi^2 f^2}.
\end{eqnarray}
The relation between $\gamma_n$ and the eigenvalues $\lambda_n$ follows from the structure of the unnormalized eigenfunction obtained from Eq.~(\ref{eq:eignPn}) and given as $P_n(\gamma) = \gamma^{-5/2}J_1(2\sqrt{2\lambda_n/\gamma})$, where $J_1(x)$ is the Bessel function of the first kind. Hence, the first moment of it is $\gamma_n = (2\lambda_n)^{-1/2}$. The regime $S(f)\sim 4\int_0^\infty \frac{d\lambda}{\lambda^2+4\pi^2f^2} = 1/f$ is obtained when $\gamma_n\sim \lambda_n^{-1/2}$.

The joint appearance of the $1/f$-fluctuations distributed with an inverse cubic law for the flux of tweets and volume of trades has interesting implications. Although, there is no consensus on a unique underlying process generating $1/f$-noise, we associate this kind of fluctuations to complex systems that exhibit an increase in structures and information due to a long-term memory. In general, the collective interactions in relation to trading and tweeting exhibit the characteristics of an emergent phenomena. 

Intermittent fluctuations in the rate of tweets on Twitter 
can happen by two types of information pathways: i) the influx of tweets triggered from the outside world onto Twitter (many independent users tweet about the same thing), ii) cascades of retweets or replies to existing tweets. Since there is a larger influx of new tweets compared to the rate of retweets, we conclude that most of the information spread across the online network happens in a ``one-step" cascade when many unrelated people tweet about the same thing. 

The activity on Twitter may not be very different from the dynamics of stock trades on financial markets since both are influenced by the social behavior inside massive communities combined with simple rules on the interaction set by e.g. the platform through which the individuals interact. Furthermore, the similarity on large scales indicate a common feature in the complex process underlying the decision-making of users on Twitter and participants in financial markets.

\subsection*{Materials}
From the Twitter Application Programming Interface (API), we automatically query for tweets on Twitter containing one or more of 92 preselected international brand names. For each query to Twitter a maximum number of 1500 tweets is returned by the API. Each returned tweet has a time stamp, which can be used to estimate an average tweet rate. The dataset used in this study was created by monitoring the public timeline for a period of four months November 2010 to February 2011, two months January--February 2012 and two months October--November 2012. During these periods, we computed with a sampling rate down to half a minute the tweet rates of selected international brands. 

\subsection*{Acknowledgments}
This study was supported by the Danish National Research Foundation through the Center for Models of Life. Suggestions and comments by Anders Blok are gratefully acknowledged.

\end{document}